\documentclass[lettersize,journal]{IEEEtran}
\usepackage{amsmath,amsfonts}
\usepackage{algorithmic}
\usepackage{algorithm}
\usepackage{array}
\usepackage{textcomp}
\usepackage{stfloats}
\usepackage{url}
\usepackage{verbatim}
\usepackage{graphicx}
\usepackage{cite}
\usepackage{subfigure}
\usepackage{balance}
\usepackage{enumitem}
\usepackage{booktabs}
\usepackage{multirow} % For multirow feature used in tables
\usepackage{marvosym}

\begin{document}
\title{DCT-MARL: A Dynamic Communication Topology  Based MARL Algorithm for Platoon Control}
\author{Yaqi Xu, Yan Shi\textsuperscript{\Letter}, Jin Tian,~\IEEEmembership{Graduate Student Member, IEEE}, Fanzeng Xia, \\ Tongxin Li,~\IEEEmembership{Member,~IEEE}, Shanzhi Chen,~\IEEEmembership{Fellow,~IEEE}, Yuming Ge

\thanks{Yaqi Xu, Yan Shi, and Jin Tian are with State Key Laboratory of Networking and Switching Technology, Beijing University of Posts and Telecommunications, Beijing 100876, China (e-mail: xuyq19@bupt.edu.cn; shiyan@bupt.edu.cn). Fanzeng Xia and Tongxin Li are with the Chinese University of Hong Kong, Shenzhen, China. Shanzhi Chen is with the National Engineering Research Center of Mobile Communications and Vehicular Networks, and State Key Laboratory of Wireless Mobile Communications, China Academy of Telecommunication Technology, Beijing, China. Yuming Ge is with the Key Laboratory of Internet of Vehicle Technical Innovation and Testing, Ministry of Industry and Information Technology, Beijing, China.
}}

\maketitle

\begin{abstract}
With the rapid advancement of vehicular communication facilities and autonomous driving technologies, connected vehicle platooning has emerged as a promising approach to improve traffic efficiency and driving safety. Reliable Vehicle-to-Vehicle (V2V) communication is critical to achieving efficient cooperative control. However, in the real-world traffic environment, V2V communication may suffer from time-varying delay and packet loss, leading to degraded control performance and even safety risks. To mitigate the adverse effects of non-ideal communication, this paper proposes a Dynamic Communication Topology based Multi-Agent Reinforcement Learning (DCT-MARL) algorithm for robust cooperative platoon control. Specifically, the state space is augmented with historical control action and delay to enhance robustness against communication delay. To mitigate the impact of packet loss, a multi-key gated communication mechanism is introduced, which dynamically adjusts the communication topology based on the correlation between vehicles and their current communication status. Simulation results demonstrate that the proposed DCT-MARL significantly outperforms state-of-the-art methods in terms of string stability and driving comfort, validating its superior robustness and effectiveness.

\end{abstract}


\section{Introduction}
\label{sec:introduction}
% BackGround
In recent years, with the rapid deployment of Vehicle-to-Vehicle (V2V) communication facilities and autonomous driving technologies, connected vehicle platooning has become a promising approach for enhancing road safety and traffic efficiency \cite{sacone2024platoon}. Through V2V communication, connected vehicles can form self-organized platoons and maintain a short inter-vehicle distance and coordinated driving behavior, thereby improving road throughput while reducing energy consumption and emissions \cite{nandhini2024comprehensive}. 

In such platooning systems, the effectiveness of cooperative control critically depends on the real-time sharing of vehicle driving state information (e.g., position, velocity, acceleration, etc.). Consequently, reliable V2V communication is an essential prerequisite for ensuring control performance \cite{balador2022survey}. However, in the real-world traffic environment, channel fading, signal obstruction, and spectral interference may induce substantial communication quality variations, resulting in time-varying delay and packet loss.

These imperfections not only hinder timely and accurate information exchange, as delay leads to outdated data and packet loss results in missing information, but also directly affect the string stability and driving comfort of the platooning system \cite{boubakri2020intra}. Hence, developing robust platoon control strategies under non-ideal communication conditions has become a pressing research challenge.

% Related work 
To address these issues, a range of delay compensation based robust control strategies have been proposed. Some studies model deterministic or random delays to improve the robustness of control algorithms \cite{zhao2020vehicle, liu2023multitimescale}. More recently, researchers have also explored solutions within the Reinforcement Learning (RL) framework by expanding the agent’s state space to include historical control actions \cite{gao2024drl, wang2022design}, which improves the system’s robustness under delayed communication. Nevertheless, these methods largely rely on simplified delay assumptions and lack explicit sensitivity to the time-varying nature of delay, limiting their scalability to highly dynamic communication environments.

For packet loss, mitigation strategies typically rely on holding previous inputs, interpolating historical data, or fallback modes \cite{ma2020distributed,liu2022integrated, bian2024distributed,navas2019mixing,zhang2024abnormal,wang2021model}. Although these methods can maintain basic control performance under mild packet loss, they often fail to guarantee stability and safety in high-loss or highly dynamic environments. Some recent studies have introduced dynamic communication topology mechanisms that adjust weights based on communication quality, aiming to mitigate performance degradation \cite{liu2022connected,wang2024self}. While this mitigates the performance degradation from communication packet loss, these strategies focus solely on communication quality metrics and overlook the need to consider the relevance of communication vehicles in the current dynamic traffic environment. In reality, the relevance of each neighboring vehicle is dynamically determined. For example, during acceleration, a driver’s attention naturally shifts to the vehicle ahead, as its motion governs the feasibility of speeding up, whereas during deceleration, the vehicle behind becomes more relevant, since its timely response is critical for avoiding rear-end collisions \cite{lappi2022gaze}.

Moreover, most existing studies typically address either communication delay or packet loss in isolation, lacking a unified control framework that considers their coupled impact. Although some traditional optimization-based works have proposed integrated modeling approaches \cite{vegamoor2021string, SHEN2024106703,wang2024stability}, they typically rely on accurate system models and analytical formulations, which limits their adaptability and generalization under real-world uncertainties and unmodeled dynamics. Therefore, it’s necessary to investigate novel intelligent algorithms to solve the adaptability problem in dynamic and complex traffic scenarios.

In contrast, Multi-Agent Reinforcement Learning (MARL), has recently emerged as a promising solution for cooperative platoon control due to its model-free nature, strong generalization capabilities, and support for distributed decision making \cite{peake2020multi, lin2024enhancing, xu2024multi}. Under assumptions of ideal communication or deterministic delay, previous studies have validated MARL’s efficacy in platoon control \cite{hua2023energy,wang2024multi}. Some methods further enhance MARL by introducing communication-aware designs, that though consider inter-agent information correlations to build dynamic communication topologies \cite{ding2020learning}. However, these approaches tend to focus on correlation among agents, while overlooking the dynamic of communication quality.  

To bridge this gap, this paper proposes a novel Dynamic Communication Topology based Multi-Agent Reinforcement Learning (DCT-MARL) algorithm to enhance the control performance of connected vehicle platoon under non-ideal communication condition. By dynamically adjusting the communication topology and augmenting the state space, the proposed method improves the robustness and adaptability of vehicle platoon in realistic communication environment. The main contributions are summarized as follows:

%Results
%Contribution

\begin{itemize}
	\item[1)]
A multi-key gated communication network is proposed,
enabling each agent to infer the relevance of communication agents based on real-time observations and historical
contexts. This mechanism dynamically adjusts the communication topology in response to packet loss.
	
	\item[2)] A state enhancement scheme is designed to mitigate
the impact of communication delay on platoon control.
Specifically, the agent’s state is extended to include both
the history control action and delay information. This
increase improves the policy sensitivity to time-varying communication delay and improves the
robustness and adaptability of the system to delayed communication.
	
	\item[3)]
A co-simulation environment integrating vehicular control and wireless communication is developed to enable
comprehensive evaluation under realistic condition. Extensive experiments demonstrate that the proposed DCT-MARL algorithm outperforms state-of-the-art methods in
terms of string stability and driving comfort under non-ideal communication condition.

\end{itemize}

% Overview

The remainder of this paper is organized as follows: Section \ref{sec:rel} introduced related work on platoon control under non-ideal communication scenarios, as well as MARL-based control approaches. Section \ref{sec:prob} formulates the problem. The details on proposed DCT-MARL based control method is discussed in Section \ref{sec:macc}. Section \ref{sec:set} presents the experimental setting and Section \ref{sec:experence} discusses the experimental results. Finally, the conclusions and future works are given in Section \ref{sec:conclusion}.

\section {Related Work}\label{sec:rel}
\subsection{Platoon Control under Unideal Communication Conditions}

Reliable V2V communication is essential for cooperative platoon control, however, packet loss and delay in practical V2V communication may induce performance degradation in the control system \cite{balador2022survey,boubakri2020intra}.

To mitigate the negative impact of communication delay, a range of delay compensation based robust control strategies have been proposed \cite{zhao2020vehicle,liu2024compensation,gao2024drl,wang2022design,liu2023multitimescale}. To address packet loss, traditional approaches employ techniques include Zero-Order Hold (ZOH) to maintain previous control inputs \cite{ma2020distributed}, historical data interpolation \cite{liu2022integrated,bian2024distributed}, or fallback modes such as Adaptive Cruise Control (ACC) \cite{navas2019mixing, zhang2024abnormal,wang2021model}. To enhance adaptability, some recent studies have introduced dynamic communication topology mechanisms that adjust weights based on communication quality, thereby mitigating the effects of communication packet loss \cite{liu2022connected,wang2024self}. For example, Liu et al. \cite{liu2022connected} developed an traditional optimization-based method to adjust weights according to communication condition, whereas Wang et al. [16] proposed a self-adaptive topology control method based on a Dual-Decision Deterministic Policy Gradient ($D^3PG$) algorithm for platoon control, which significantly enhances system resilience under fluctuating communication condition. However, these strategies primarily focus on communication quality metrics, overlooking the fact that in practical applications, the selection of communication targets should be context-aware and responsive to dynamic traffic conditions.

Furthermore, most existing studies address either delay or packet loss in isolation, making them insufficient for scenarios where both impairments coexist. In real-world environments, however, delay and packet loss frequently occur simultaneously, creating compounded challenges for platoon control systems. To address this, some recent studies have proposed unified models that simultaneously account for the effects of delay and packet loss \cite{zhao2020vehicle,ma2020distributed,vegamoor2021string,SHEN2024106703,wang2024stability}. For instance, Vegamoor et al. \cite{vegamoor2021string} modeled delay and packet loss simultaneously, and introduced a safety-aware constraint on distance errors between any two vehicles and derived a lower bound on safe time gaps to ensure string stability under unideal communication scenarios. Zhao et al. \cite{zhao2020vehicle} examined control strategies under constant time headway distance, modeling non-ideal communication conditions including limited communication range, random packet loss, and time-varying delay. Their approach employed a distance-based dynamic topology and a robust controller that utilized multi-leader an follower information. Following this, Wang et al. \cite{wang2024stability} further advanced the approach by proposing a platoon following model based on dynamic weight optimization, which adaptively adjusted the weighting of incoming information. Their approach integrated realistic vehicle dynamics and demonstrated strong robustness under non-ideal communication conditions.

While these control approaches provide valuable theoretical foundations, they often rely heavily on accurate system modeling and incur significant computational overhead. Consequently, they are difficult to scale in dynamic and uncertain traffic environment characterized by rapidly changing velocities and nonlinear interactions.

\subsection{MARL-based Platoon Control}
In recent years, MARL has gained significant attention in intelligent transportation systems due to its strengths in distributed decision-making, policy learning, and environmental adaptability\cite{hua2023energy, chen2024communication, parvini2023aoi,wang2024multi}. MARL has become a promising paradigm for cooperative control tasks, such as vehicle platooning. For example, Chen et al.\cite{chen2024communication} evaluated MARL frameworks using both local and global reward structures in platoons of three and five vehicles, and found that decentralized learning with local rewards achieved superior performance in terms of stability and responsiveness compared to centralized approaches.

However, the vast majority of existing MARL-based platoon control studies assume ideal V2V communication, free of delay and packet loss, and thus lack a comprehensive evaluation into the robustness of learned policies under realistic communication impairments. To address this gap, Wang et al. \cite{wang2024multi} proposed a delay-aware MARL framework based on a partially observable Markov game (POMG), capturing delay-induced uncertainties in platoon systems. Their simulations demonstrated improved platoon stability under time-varying delays. However, existing works still fall short of simultaneously addressing communication delays and packet losses in a unified MARL framework.

On dynamic communication topologies within MARL, researchers have explored inter-agent information correlations to construct adaptive communication topologies. Seminal algorithms including ITGCNet\cite{zhang2025bridging}, Gated-ACML\cite{mao2020learning}, I2C\cite{ding2020learning}, and NeurComm\cite{chu2020multi} employ gating or attention mechanisms to determine when and with whom to communicate, thereby improving learning efficiency. In particular, Chu et al.\cite{chu2020multi} proposed a learnable protocol that enables agents to formulate distributed policies based on both local and neighbor information, partially mitigating the adverse effects of communication failure. Ding et al. \cite{ding2020learning} propose I2C, a dynamic communication framework where each agent learns to infer its communication targets from local observations via a prior network trained through causal inference. However, these methods primarily emphasize interagent information relevance while neglecting the dynamic variability of the communication environment, limiting their applicability under real-world communication environment.  

To overcome these limitations, this paper proposes a DCT-MARL framework that jointly addresses communication delay and packet loss. The framework aims to substantially enhance the robustness and adaptability of connected vehicle platoon in non-ideal communication environment.

\section{Problem Formulation}\label{sec:prob}

\subsection{Vehicle Dynamics Model}\label{sec:vdm}

This paper considers a platoon consisting of $N+1$ vehicles, indexed as $0,1,…,n$, where the leading vehicle is indexed as $0$, and the following vehicles are indexed from $1$ to $n$. All vehicles communicate through V2V wireless communication to exchange information and coordinate movements. In this study, this paper primarily focus on a linear model of a connected automated vehicle (CAV) system, which has been extensively used in vehicle control. The vehicle dynamics are given by \cite{chu2019model}.

\begin{equation}
	\begin{gathered}
		p_{i, t+1}=p_{i, t}+\Delta t * v_{i, t} \\
		v_{i, t+1}=v_{i, t}+\Delta t * a c c_{i, t} \\
		a c c_{i, t+1}=\left(1-\frac{\Delta t}{\tau_i }\right) a c c_{i, t}+\frac{T}{\tau_i} * u_{i, t},
	\end{gathered}
\end{equation}
where $p_{i, t+1}, v_{i, t+1}$ and $a c c_{i, t+1}$ represents the position, velocity, and acceleration of vehicle $i$ at time step $t+1$, respectively. The $\Delta t$ is the sampling time. The term $u_{i, t}$denotes the control input (i.e., commanded acceleration), and $\tau_i $ is a time constant representing driveline dynamics, describing how the acceleration responds to the control input. 

In this study, this paper adopt the constant time headway policy (CTHP), meaning that the platoon maintains a constant time gap between adjacent vehicles. Each following vehicle $i$ aims to maintain a desired headway distance from its preceding vehicle $i-1$, which is defined as the bumper-to-bumper distance. The ideal inter-vehicle distance between vehicle $i$ and its predecessor $i-1$ is denoted as 
$d_{r,i,t}$.
\begin{equation}
	d_{r, i, t}=d_{0}+h * v_{i, t},
\end{equation}
where $h>0$ is the constant time headway, and $d_{0}$ is the standstill distance that ensures safety when vehicles are stopped.
The actual inter-vehicle distance between vehicle $i$ and vehicle $i-1$ is given by $ d_{i, t}=p_{i-1,t}-p_{i,t}-l$, where $l$ is the vehicle length. The distance error for vehicle $i$ is then computed as:
\begin{equation}
	e r r_{p_i, t}=d_{i, t}-d_{r, i, t}=p_{j, t}-p_{i, t}-l-d_{0}-h * v_{i, t}.
\end{equation}

Since all vehicles in the platoon are assumed to be identical, $d_{0}$ and $l$ are considered uniform across all vehicles and are therefore omitted in the subsequent derivations.

For any adjacent vehicle pair $(i,j)$ in the communication graph, the tracking error consists of both distance error and velocity error, defined as follows:
\begin{equation}
	\label{equ:err}
	\begin{aligned}
		\mathrm{err}_{p_{i,t}} &= 
		\begin{cases}
			p_{j,t} - p_{i,t} - \sum_{\mu = j+1}^{i} h \cdot v_{\mu,t}, & i > j \\
			p_{i,t} - p_{j,t} - \sum_{\mu = i+1}^{j} h \cdot v_{\mu,t}, & j > i
		\end{cases} \\
		\mathrm{err}_{v_{i,t}} &= 
		\begin{cases}
			v_{j,t} - v_{i,t}, & i > j \\
			v_{i,t} - v_{j,t}, & j > i
		\end{cases}
	\end{aligned}
\end{equation}
Here, $j < i$ means vehicle $j$ is ahead of vehicle $i$. The distance error $\text{err}_{p,i,t}$ captures the deviation between the actual distance of vehicles $i$ and $j$ and the cumulative desired distance, which is obtained by summing the product of the desired time headway $h$ and the instantaneous velocities of all intermediate vehicles between $i$ and $j$. The velocity error $\text{err}_{v,i,t}$ reflects the relative speed difference between the two vehicles. 

Each vehicle $i$ is capable of communicating with other vehicles within its communication range. However, due to potential communication disruptions or packet loss, the communication agent may become unavailable at certain time steps. Additionally, due to network congestion, signal attenuation, or interference, communication may experience a delay $\xi_{i,j}$, meaning that vehicle $i$ at time $t$ receives information from vehicle $j$ that corresponds to time $t - \xi_{i,j}$ instead of real-time data.

\subsection{Control Objectives of Vehicle Platoon}\label{sec:cm}

The primary control objective of the connected vehicle platoon is to minimize the deviation of each follower vehicle from its desired relative position and velocity with respect to its preceding vehicles even under non-ideal V2V communication conditions including delay and packet loss. Specifically, the system aims to minimize both the inter-vehicle distance error and the velocity control error, thereby ensuring safe, stable, and efficient following behavior under communication constraints.

Considering the vehicle dynamics described in Eq.~\ref{equ:err}, the control objective is to minimize the cumulative tracking error over a finite time horizon $T$:

\begin{equation}
	\min_{u_{i,t}} \sum_{t=0}^{T} \sum_{i=1}^{N} \text{err}_{i,t},
\end{equation}

where the instantaneous tracking error for vehicle $i$ at time $t$ is defined as:
\begin{equation}
	\text{err}_{i,t} = \lambda_p \|e_{p_i,t}\|^2 + \lambda_v \|e_{v_i,t}\|^2,
\end{equation}
and $\lambda_p$ and $\lambda_v$ are weighting coefficients that determine the relative importance of position and velocity tracking errors. These coefficients are typically set to equal values unless a specific prioritization is needed.

In addition, each vehicle’s behavior must satisfy physical and safety-related constraints:

\begin{equation}
	\begin{gathered}
		\mathrm{acc}_{\min} \leq \mathrm{acc}_{i,t} \leq \mathrm{acc}_{\max},
	\end{gathered}
\end{equation}
where $\mathrm{acc}_{i,t}$ is the actual acceleration of vehicle $i$. The bounds $\mathrm{acc}{\min}$ and $\mathrm{acc}{\max}$ represent the allowable acceleration range, while $u_{\min}$ and $u_{\max}$ denote the bounds on control input. These constraint ensure safe and comfortable driving and prevent excessive control efforts.

\section{DCT-MARL Based Algorithm} \label{sec:macc}
\begin{figure*}[htbp]
	\centering
	\includegraphics[width=0.9\textwidth]{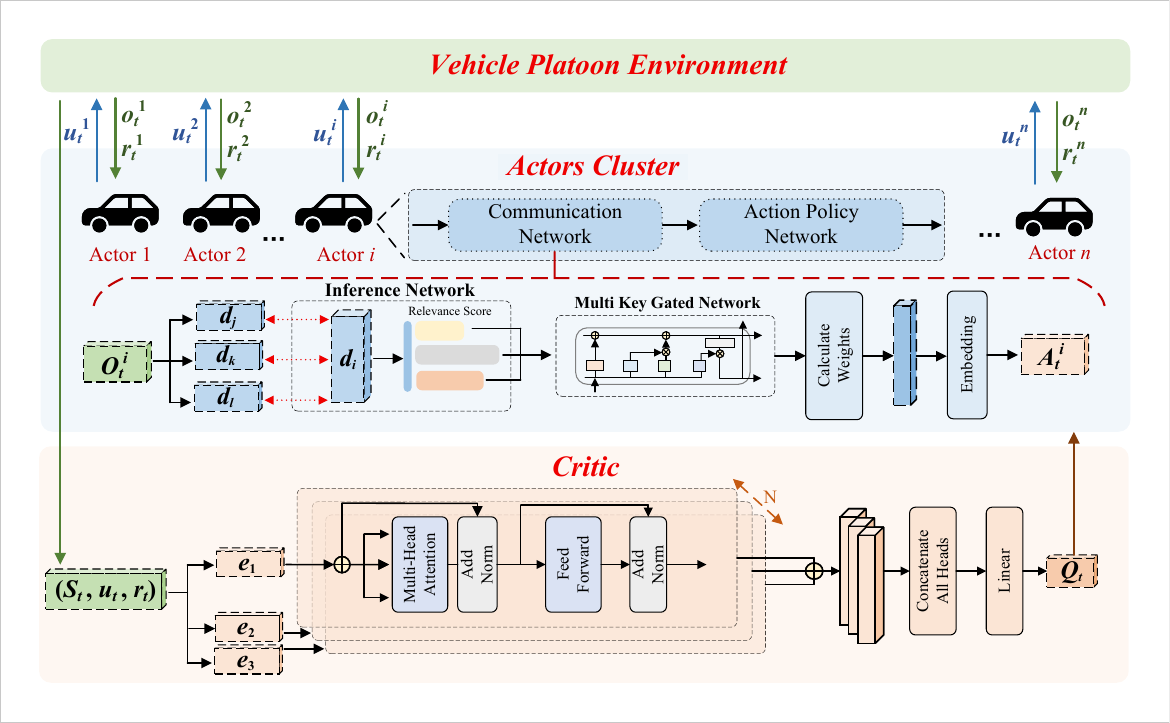}
	\caption{The architecture of DCT-MARL based Platoon Control algorithm.}
	\label{fig:DCT}
\end{figure*}

In this section, the proposed DCT-MARL algorithm is described, as shown in Figure.\ref{fig:DCT}. The platoon control task is then formulated as a Decentralized Partially Observable Markov Decision Process (Dec-POMDP). To ensure robust control performance under non-ideal communication conditions, the framework integrates two key mechanisms: (i) a dynamic communication topology that adaptively selects communication partners, and (ii) an explicit delay module that augments the state and reward design to account for time-varying transmission latency. 

The DCT-MARL algorithm is built upon the actor-critic architecture. Each actor consists of a multi-key gated communication network, which dynamically selects relevant neighbors based on inference mechanisms, and an action policy network that outputs control actions. At each control step, every agent gathers state information from vehicles within its communication range. Considering the possibility of communication failures, the agent adaptively determines its communication topology based on local observations, integrating self-state and received information for decision-making.

This design is well aligned with real-world vehicle platoon systems, where the communication topology is frequently affected by wireless interference, signal blocking, and vehicle mobility. The dynamic topology selection mechanism allows each vehicle to adaptively adjust its communication partners, ensuring that critical information is exchanged even under non-ideal communication conditions.

The centralized critic aggregates joint actions and state representations from all agents to estimate each agent’s actionvalue function at each time step. An attention mechanism is utilized to extract relevant information from other agents’  observation-action pairs, enabling each agent to selectively filter critical environmental information. The critic network is trained via temporal difference learning \cite{du2021learning} and is only used during training, enabling decentralized execution once the actors are deployed.

To optimize each agent $i$'s policy $\pi_i$, this paper minimizes the Kullback--Leibler (KL) divergence between the learned distribution $q_{\theta_i}$ and the target distribution $p$:

\begin{align}
	-D_{\text{KL}}(q_{\theta_i} \parallel p) = \sum_{t} \mathbb{E}_{s_t, u_t, A_t \sim q_{\theta_i}} \Big[ 
	& r_i(s_t, u_t) + H\big( \pi_i (u_i \mid o_i(s, A)) \nonumber \\
	& \cdot \rho(A_t \mid s_t) \big)
	\Big].
\end{align}

Beyond the reward term, the objective introduces an additional conditional entropy term for the joint distribution 
$\pi_i(a_i \mid o(s_t, A_t)) \, \rho(A_t \mid s_t)$, 
which encourages exploration for agent $i$ and enhances the graph policy $\rho$. The adjacency trajectory matrices for the dynamic directed graph and its subgraphs are denoted as $\mathbf{A}$ and will be further detailed in the following section. The objective function for agent $i$ is thus redefined as:

\begin{equation}
	J_{\pi_i}(\theta_i) = \mathbb{E}_{q_{\theta_i}} \left[ Q(s, u) + H\left( \pi_i (u_i | o_i(s, A)) \rho(A | s) \right) \right].
\end{equation}

The policy $\pi_i$ is updated by maximizing $J_{\pi_i}$, and its gradient is:

\begin{align}
	\nabla_{\theta_i} J(\theta_i) = \mathbb{E}_{q_{\theta_i}} \Big[
	& \nabla_{\theta_i} \log \pi_i(u_i \mid o_i(s, A)) \cdot \nonumber \\
	& \left( -\alpha \log \pi_{\theta_i}(u_i \mid o_i(s, A)) + Q_i(s, u) \right)
	\Big].
\end{align}

The critic network $Q_i$ is trained by minimizing the temporal-difference loss:

\begin{equation}
	L_Q (\phi) = \mathbb{E}_{(s_t, u_t) \sim D} \left[ (y_t^i - Q_i (s_t, u_t; \phi))^2 \right],
\end{equation}

where the TD target is:

\begin{equation}
	y_t^i = r_i (s_t, u_t) + Q_i (s_{t+1}, u_{t+1}^i, u_{t+1}^{-i}; \phi).
\end{equation}

During training, this paper follow these iterative procedures:
\begin{itemize}
	\item First, fix $\pi_i (u_i | s, A)$ and $Q_i (s, u)$ to update $\rho(A)$.
	\item Then, improve $\pi_i (u_i | o(s, A))$ and $Q_i (s, u, A)$ using the generated trajectories given $A$.
\end{itemize}

The prior network, parameterized by $ \theta_{b_i}$, is trained as a binary classifier using the training dataset $S$ and is updated with the following loss function:

\begin{align}
	\mathcal{L}(\theta_{b_i}) = \mathbb{E}_{(o_i, d_j), \mathcal{I}_i^j \sim \mathcal{S}} \Big[ 
	& -(1 - y_i^j) \log \left(1 - b_i(o_i, d_j)\right) \nonumber \\
	& + y_i^j \log \left(b_i(o_i, d_j)\right) 
	\Big],
\end{align}

where the binary label defined as:
$
y_i =
\begin{cases} 
	1, & \text{if } P_{\omega}(y_i \mid o_i) \geq \delta \\
	0, & \text{otherwise},
\end{cases}
$
where $\delta$ is a tunable hyperparameter controlling the confidence threshold for classification. Notably, if a certain agent significantly impacts another agent’s state estimation during centralized training, its information must remain accessible during decentralized execution. To achieve this, the dynamic communication topology learned in training is retained for execution, enabling consistent and robust coordination.

\subsection{Dec-POMDP Formulation}

A Dec-POMDP is formally defined as a tuple $\langle N, S, U, P, R, O, \gamma \rangle$, where $N = {1, \dots, n}$ denotes a finite set of $n$ agents interacting in a shared environment. The environment is characterized by a global state space $S$, and each agent selects actions from a joint action space $U$. The environment dynamics are governed by the state transition function $P: S \times U \rightarrow S$, which defines the probability distribution over subsequent states given the current state and joint actions. Each agent receives observations from an observation space $O$, which may provide only partial and possibly noisy information about the true state. The reward function $R: S \times U \rightarrow \mathbb{R}$ maps the current state and joint actions to a scalar reward shared among all agents. Finally, $\gamma \in [0, 1)$ is the discount factor that determines the present value of future rewards.

At each time step $t$, each agent $i \in N$ receives an observation from the state and follows an individual policy $\pi_i$, where $\pi_i$ depends on the agent's observation-action history. The joint action of all agents leads to the next state, and the team receives a global reward. Each agent aims to maximize the accumulated reward.

State Space $S$: To mitigate the performance degradation caused by communication delays, this paper extend the definition of the agent’s state space by incorporating both historical control information and delay descriptors. Specifically, at time step $t$, the state of vehicle $i$ is defined as:
\begin{equation}
	S_{i,t} = \left\{p_{i,t}, v_{i,t}, acc_{i,t}, u_{t-1}, \xi_{i,t} \right\},
\end{equation}
where $p_{i,t}$, $v_{i,t}$, and $acc_{i,t}$ denote the vehicle’s current position, velocity, and acceleration, respectively; $u_{t-1}$ represents the most recent control input, and $\xi_{i,t}$ denotes the communication delay associated with the latest received information. 

Although this formulation deviates from the conventional state definition, it is a rational extension under communication-impaired environments, where both the delayed control input and latency should be explicitly considered in the state representation. In connected vehicle systems, time-varying wireless communication often causes neighboring information to arrive with variable delays. If the control policy relies solely on such delayed observations, the Markov property may be compromised, thereby impairing decision-making performance.

By incorporating the previous control action $u_{t-1}$, the agent can better infer the underlying system dynamics during periods of delayed or missing observations. Additionally, explicitly modeling the delay $\xi_{i,t}$ enables the policy to assess the reliability of received data and adjust its decisions accordingly. Upon receiving information from neighboring agents, the agent estimates the communication delay using the attached timestamps and incorporates this value into its current state. This augmented state representation preserves the Markovian assumption under delayed observations and significantly enhances decision-making robustness in non-ideal communication environments.

Observation Space $O$: Agents use communication to obtain information beyond their direct observations to assist decision-making. Each agent’s local observation $o_i \in S_i$ is a \textit{partial description} of the environment state. At time step $t$, each agent $j$ in the platoon transmits its driving information$ X_{j,t} = {p_{j,t}, v_{j,t}, acc_{j,t}, t_{j,t}}$, neighboring agent $i$ obtains driving information from multiple vehicles within its V2V communication range, forming its \textit{partial observation} $o_i$ and determining which agents are within communication range. This solves the problem of possible communication failures. For example, if agent l fails within the current control interval k, agent i cannot observe agent l's information through V2V communication, and can receive the status information of other related vehicles for control output. It avoids the problem that the amount of states received may be insufficient due to communication failures. The global state $O$ is then defined as the combination of all agents' local observations:$ O = \{ o_1, o_2, \dots, o_n \}.$
Similarly, the global state $S$ is defined as: $S = \{ s_1, s_2, \dots, s_n \}.$

Action Space $U$: Each agent $i$ follows a decentralized policy 
$
\pi_i: O_i \times S_i \to [0,1],
$
to determine the control output at time step $t$. Each PC agent $i$ selects a control input as its PC action at  time step $t$.

Reward Function $R$: The reward function $r_{i,t}$ is critical for training RL agents to exhibit the desired behavior. Our objective is to train agents to ensure platoon stability and comfort under imperfect communication and dynamic traffic conditions. The reward assigned to agent $i$ at each time step $t$ is designed as follows:

\begin{align}
	r_{i,k} = -\big\{ &
	\omega_1\left(\text{err}_{p_i, k}\right)
	+ \omega_2\left(\text{err}_{v_i, k}\right)
	+ \omega_3\left(\text{jerk}_i\right) \\
	& + \omega_4 \cdot \operatorname{ReLU}\left(d_{\text{safe}} - d_{i,k}\right)
	\big\},
\end{align}

where $w_1$, $w_2$, $w_3$, and $w_4$ are the weighting coefficients. In this equation, the first two terms penalize deviations from ideal position and velocity, encouraging agents to achieve these goals closely and guaranteeing the stability of the platoon. The third term is to improve driving comfort to minimize sudden acceleration, thereby providing a smoother and more comfortable ride for passengers. Where jerk is the rate of change of acceleration, which is given by
\begin{equation}
\text{jerk}_i=-\frac{\dot{acc}_i^2(t)}{\left[\frac{\left(u_{\max }-u_{\min }\right)}{\Delta t}\right]^2} .
\end{equation}

The fourth item acts as a safety constraint, severely penalizing agents if the distance between vehicles is less than the safe distance, which is essential to prevent collisions and ensure the safety of the vehicle queue. $d_{i,k}$ is the actual distance between vehicle $i$ and the vehicle in front, and $d_{\text{safe}}$ is the preset minimum safe distance. $\operatorname{ReLU}(\mathrm{x})=\max (0, x)$ used to ensure that a safety item generates a penalty only if the actual distance is below the safety threshold.

In addition, considering the communication delay problem, each agent can only access the lagged observation information when making a decision, and cannot directly obtain the current true state. Therefore, this paper modify the reward function based on conditional expectation to adapt it to the optimal learning in the communication constrained environment. Specifically, for each agent $i$, in the extended state $S_{i, t}$, the reward function of delay is defined as follows:

\begin{equation}
	\tilde{R}_{i, t}\left(S_{i, t}, acc_{i, k}^{\mathrm{CL}}\right) = \mathbb{E}_{x_{i, t}}\left[R_{i, t}\left(x_{i, t}, acc_{i, t}\right) \mid S_{i, t}\right].
\end{equation}

This modification ensures that even in a communication-constrained environment, agents can still make reasonable decisions based on past observations and actions while maximizing their long-term rewards.

Therefore, each agent determines its action based on its own observations and the messages received in a distributed manner. This process is repeated for each time step under different observations. This work aims to achieve comfort and safety control of platoon. The ultimate goal of the RL model is:

\begin{equation}
	\max \left(J_\pi(s(0))=\lim _{M \rightarrow \infty} \mathbb{E}\left[\sum_{t=0}^{M-1} \gamma^k r_i(s(t), \pi(s(t)))\right]\right) ,
\end{equation}
where $J_\pi(s(0)$ is the expected cost with initial condition $s(0)$ and control policy $\pi$, $\ \gamma$ is the discounted factor, $M$ is the number of steps; $E(\cdot)$ is the mathematical expectation. The update strategy of the proposed DCT-MARL is given in Algorithm 1.
\begin{algorithm}[h]
	\caption{DCT‑MARL: Dynamic Communication Topology‑based Multi‑Agent RL.}
	\label{alg:optimization_algorithm}
	\textbf{Ensure:} Action Policy $\pi_i$, and graph reasoning policy $\rho$ \\
	\textbf{1:} Initialize parameters $\theta_i$, $\phi_i$ for each agent and $\phi$ for graph reasoning; \\
	\textbf{2:} Assign target parameters of Q-function: $\phi_i \leftarrow \phi_i$, and target policy parameter: $\theta_i \leftarrow \theta_i$ \\
	\textbf{3:} $D \leftarrow$ empty replay buffer \\
	\textbf{4:} \textbf{while} not terminated \textbf{do} \\
	\textbf{5:} \hspace{0.5cm} Sample $A \sim \rho_{\phi}(A | s)$; \\
	\textbf{6:} \hspace{0.5cm} Given current state $s$, compute $o_i(s, A)$ for all agents and $x$; \\
	\textbf{7:} \hspace{0.5cm} Sample $u_i \sim \pi_i(\cdots | o_i(s, A))$; \\
	\textbf{8:} \hspace{0.5cm} Take the joint action $\{u_i\}_{i \in N}$ and observe own reward $r_i$ and new state $s'$: \\
	\hspace{0.5cm} $D \leftarrow D \cup \{(s, \{u_i\}_{i \in N}, A, \{r_i\}_{i \in N}, s')\}$; \\
	\textbf{9:} \hspace{0.5cm} \textbf{for} each agent \textbf{do} \\
	\textbf{10:} \hspace{1cm} Update $\phi_i$ according to $\nabla_{\phi} J_Q(\phi_i)$; \\
	\textbf{11:} \hspace{1cm} Update $\theta_i$ according to $\nabla_{\theta_i} J_{\pi}(\theta_i)$; \\
	\textbf{12:} \hspace{1cm} Update $\phi$ according to $\frac{1}{n} \sum_{i=1}^n \nabla_{\phi} \eta_i(\rho_\phi)$; \\
	\textbf{13:} \hspace{1cm} Reset target Q-function: $\phi_i = \beta \phi_i + (1 - \beta) \phi_i$; \\
	\textbf{14:} \hspace{0.5cm} \textbf{end for} \\
	\textbf{15:} \textbf{end while}
\end{algorithm}

\subsection{Dec-POMDP with Dynamic Communication topology}

The vehicles in the platoon exchange their state information (such as position, velocity, and acceleration) through a V2V wireless communication network. Each vehicle has its own communication network, which can be modeled as a directed weighted graph $G=(T,V,E)$, where $t \in T$ is the set of the trajectory in the time dimension, $V$ represents the set of vehicles. $E \subseteq V \times V$ represents the set of communication links, where $ (i,j) \in E$ indicates that vehicle $i$ can send information to vehicle $j$, representing the communication connection between vehicles. The adjacency trajectory matrices for the dynamic directed graph and subgraphs are denoted by $\mathbf{A}$. When a node can access the observations of other nodes, it can approximate the global state.

Based on Dec-POMDP, this paper propose a communication topology using dynamic directed graphs. The structural properties of the graph are represented by the adjacency trajectory matrix $A$, which is denoted as $\mathbf{A} \in \mathbb{B}^{l \times n \times n}$, where B indicates the Boolean matrix, l denotes the length of the trajectory and n indicates the number of agents. If $\lceil A^{ij}_t \rceil = 0$, then no communication occurred from agent $i$ to agent $j$ at time $t$. If $\lceil A^{ij}_t \rceil = 1$, then communication transpired from agent $i$ to agent $j$ at time $t$.

\subsection{Multi-Key Gating Network-Based Dynamic Communication topology}
During communication, each agent intuitively prefers to interact with those agents that significantly influence its decision-making strategy. Specifically, assuming agent $i$ can receive information from agent $j$, it first inputs its local observation $o_i$ along with the ID of agent $j$ into the model, which then outputs a confidence score indicating the likelihood of communication with agent $j$. 

To quantify this causal influence, a feedforward neural network is employed to implement causal inference, leveraging the KL divergence to measure the causal confidence between agent $i$ and other agents:

\begin{equation}
    D_{\mathrm{KL}}\!\left( P(u_i \mid u_{-ij}, o) \,\big\|\, 
    P(u_i, u_j \mid u_{-ij}, o) \right),
    \label{eq:causal_kl}
\end{equation}

where $P(u_i \mid u_{-ij}, o)$ and $P(u_i, u_j \mid u_{-ij}, o)$ represent two conditional probability distributions over the action space of agent $i$, and $o$ denotes the joint observations of all agents. Here, $u_{-i}$ refers to the joint actions of all agents excluding agent $i$, and $u_{-ij}$ denotes the joint actions of all agents excluding both agents $i$ and $j$.

The latter probability distribution does not depend on agent $j$'s actions, meaning that when making decisions, agent $i$ ignores the actions of agent $j$. Note that the conditional actions of other agents $u_{-i}$ are sampled from their current policies. The magnitude of KL divergence quantifies how agent $i$ would adjust its own policy if it considered agent $j$’s actions, thus indicating the degree of correlation between agent $j$’s policy and agent $i$’s policy. 

These probability distributions are computed via the joint action-value function. For instance, the softmax distribution over agent $i$’s actions is given by:

\begin{equation}
    P(u_i \mid u_{-ij}, o) = 
    \frac{\exp\!\left(Q(u_i, u_{-ij}, o)/\lambda\right)}
         {\sum_{u_i'} \exp\!\left(Q(u_i', u_{-ij}, o)/\lambda\right)},
    \label{eq:softmax}
\end{equation}

where $Q(u, o)$ is the joint action-value function, and 
$\lambda \in \mathbb{R}^+$ is the temperature parameter. 
The marginal distribution $P(u_i \mid u_{-ij}, o)$ is computed as follows:

\begin{equation}
    P(u_i \mid u_{-ij}, o) 
    = \sum_{u_j} P(u_i, u_j \mid u_{-ij}, o).
    \label{eq:marginal}
\end{equation}

After computing causal influence, a multi-key gating network is used to select $m$ communication agents. Using the Gumbel--Softmax activation, the network determines whether an agent communicates or not:

\begin{equation}
    \text{key}_i^t = \mathrm{gumbel\text{-}softmax}\!\left(
        W_v \tanh\!\left(W_q o_i^{t-1} + W_k o_i^{t-1}\right)\right),
    \label{eq:gumbel}
\end{equation}

where Gumbel-Softmax~\cite{jang2016categorical,maddison2016concrete} is a binary activation (communicate or not), and the weight matrices $W_v$, $W_q$, and $W_k$ encode the communication relevance under dynamically changing platoon topologies. 

To improve interpretability and embed safety considerations, a distance weighting module assigns higher weights to closer vehicles, highlighting their critical role in collision avoidance:

\begin{equation}
    \omega_{j,t} = \frac{\frac{1}{2^{|i-j|}}}
    {\sum\limits_{l \in \mathcal{N}_{i,t}} \frac{1}{2^{|i-l|}}},
    \label{eq:distance_weight}
\end{equation}

where $d_{ij}$ is the distance between vehicles $i$ and $j$, and $\varepsilon$ is a small positive constant preventing division by zero. 

Using the outputs of the gating network, the adjacency matrix $\mathbf{A}$ is then constructed by aggregating keys:

\begin{equation}
    A_{i,k} = \max \big( \text{key}_i^t \big),
    \label{eq:adjacency}
\end{equation}

where $\text{key}_i^k$ denotes the number of keys assigned to agent $i$ at time step $t$, and $\lceil \cdot \rceil$ is the ceiling function. This process dynamically adjusts the communication topology, while the limited number of keys ensures that each agent only exchanges information with selected neighbors, keeping the input dimension stable. Finally, all received messages are passed through the message encoder network $e_i(\cdot)$ to obtain encoded messages $c_i$, which are fed into the policy network to generate the action distribution $\pi_i(u_i \mid c_i, o_i)$.

\section{Experimental Setting and Metrics} \label{sec:set}

\subsection{Experimental Setup}

To validate the effectiveness of the proposed control strategy, a platoon of six vehicles is modeled within a 1-km highway segment in SUMO. The leading vehicle’s driving trajectory is derived from real-world traffic data collected on the eastbound I-80 highway in the San Francisco Bay Area, sourced from the NGSIM dataset \cite{NG2007}. As a widely recognized benchmark in platoon control research, the NGSIM dataset encompasses congested traffic scenarios, enabling a rigorous evaluation of the controller’s ability to manage stop-and-go traffic oscillations frequently encountered in real-world platoon operations. Integrating such realistic driving patterns facilitates more effective policy learning and enhances the applicability of the proposed approach to connected vehicle platoons operating under complex and dynamic conditions. 

For the communication simulation, the wireless channel model incorporates both small-scale and large-scale fading effects. Small-scale fading is modeled as Rayleigh fading, which captures the rapid signal fluctuations caused by multipath propagation. Large-scale fading consists of path loss (PL) and shadow fading (SF), jointly modeling the attenuation of signal strength over distance and the slow variations caused by obstacles in the environment.

Path loss is modeled using the WINNER+ B1 model recommended by 3GPP~\cite{3gpp_tr36885}, which is defined as:  

\begin{equation}
    PL(d) = A_0 + 10n \log_{10}\!\left(\frac{d}{d_0}\right) + X_\sigma, 
    \quad \text{(dB)}
    \label{eq:pathloss}
\end{equation}

where $d_0$ is the reference distance; $A_0$ is the intercept representing the path loss at the reference distance $d_0$; $n$ is the path loss exponent, which varies for different traffic scenarios and is set to $n = 2.15$ for highway V2V scenarios; $X_\sigma$ represents shadow fading, modeled as a Gaussian random variable with zero mean and variance $\sigma^2$, caused by obstacles such as trees, buildings, and vehicles. Formally, 

\begin{equation}
    X_\sigma \sim \mathcal{N}(0, \sigma^2),
    \qquad \sigma = 3 \ \text{dB}.
\end{equation}

The simulation environment models the communication scenario and control 
algorithms using Python, and integrates with SUMO for co-simulation and 
optimization via the TraCI interface. The simulation time step and total 
duration are synchronized with SUMO to ensure consistency. Table~\ref{tab:sim_params} summarizes the key simulation parameters.

\begin{table}[htbp]
	\centering
	\caption{Simulation Parameters for Experimental Setup.}
	\label{tab:spar}
	\begin{tabular}{@{}l l@{}}
		\toprule
		Parameters                               & Value                            \\
		\midrule
            Highway Length& 1000m\\
            
            Minimum Vehicle Velocity& 0 m/s \\
            Maximum Vehicle Velocity& 30 m/s \\
            Acceleration Range& -3 m/s2 to 3 m/s2 \\
            Platoon Vehicle Number& 6\\
            
		Transmission Power                       & 23 dBm                           \\
		Number of Subchannels                    & 4                              \\
		Subchannel Resource Blocks (RBs)         & 10                               \\
		Channel Bandwidth                        & 10 MHz (50 RBs)                  \\
		Packet Size                           & 190 bytes                              \\
		Message Transmission Interval            & 100 ms                           \\
		\bottomrule
	\end{tabular}
\end{table}

In the reward function of the DCT-MARL algorithm (Equation 15), the weighting coefficients $\omega_1$, $\omega_2$, $\omega_3$, and $\omega_4$ are set to 1.0, 1.0, 0.1, and 10.0, respectively, to emphasize the penalty for unsafe following distances. The algorithm adopts a centralized training and decentralized execution framework. Both the centralized critic and policy networks are implemented with three fully connected layers, while the prior network consists of two fully connected layers. Each hidden layer contains 128 neurons. The activation function used in all networks is the Rectified Linear Unit (ReLU), defined as $\operatorname{ReLU}(x) = \max(0, x)$.

Training is conducted for 1 million steps in a simulated traffic environment, with an episode length of $600$ steps. The discount factor is set to $\gamma = 0.99$, and the learning rates for the actor and critic networks are $3.0 \times 10^{-4}$ and $2.5 \times 10^{-4}$, respectively. Each algorithm is trained three times using different random seeds to ensure robustness and generalization, and all hyperparameters are selected through a grid search process.

\subsection{Evaluation Metrics}
To comprehensively evaluate the performance of the proposed platoon control strategy, this paper adopted two categories of metrics: platoon stability and driving comfort. Platoon stability reflects the system’s ability to suppress inter-vehicle disturbances and avoid amplification of oscillations downstream. Following\cite{peake2020multi}, this paper define the stability index $S_i$ for each vehicle as the time-averaged weighted squared deviation in headway and relative velocity:

\begin{equation}
	S_i = \frac{1}{N_s} \sum_{t=1}^{N_s} s_{i,cf}(t) C s_{i,cf}^T(t),
\end{equation}
where $s_{i,cf}(t) = \left[\Delta h_{i, i-1}(t), \Delta v_{i, i-1}(t)\right]$ denotes the deviation vector between vehicle $i$ and its preceding vehicle at time $t$, $C$ is a weighting matrix, and $N_s$ is the total number of simulation steps. Lower values of $S_i$ correspond to better string stability, reflecting reduced propagation of distance and velocity errors across the platoon.

Driving comfort is assessed based on the smoothness of acceleration changes. The normalized comfort score $G_i'$ is defined as:

\begin{equation}
	G_i' = 1 - \frac{1}{N_s} \sum_{t=1}^{N_s} \left(\frac{acc_i(t) - acc_i(t-1)}{acc_{\text{max}}}\right)^2,
\end{equation}
where $acc_i(t)$ denotes the acceleration of vehicle $i$ at time step $t$, and $a_{\text{max}}$ is the maximum allowable acceleration change used for normalization. Higher values of $G_i'$ correspond to smoother acceleration transitions and thus greater driving comfort.

\section{Performance Results and Analysis} \label{sec:experence}
\subsection{Control Performance against Existing Platoon Control Strategies}

To verify the effectiveness and robustness of the proposed strategy, this paper conducts a comprehensive comparison between the proposed DCT-MARL method and other state-of-the-art methods. These include the traditional optimization-based Dynamic Weights Optimization Control (DWOC) strategy~\cite{wang2024stability}, which accounts for both communication delay and packet loss by dynamically compensating front-vehicle information through communication-quality-aware weighting, and the DRL-based $D^3$PG method~\cite{wang2024self}, which introduces a self-adaptive topology control approach that significantly enhances resilience under fluctuating communication conditions. To ensure a fair comparison, the state space settings of D3PG are aligned with our proposed method, including the incorporation of a delay compensation mechanism. 

\begin{figure*}[htbp]
	\centering
	\subfigure[DWOC Policy]{
	\includegraphics[width=0.850\textwidth]{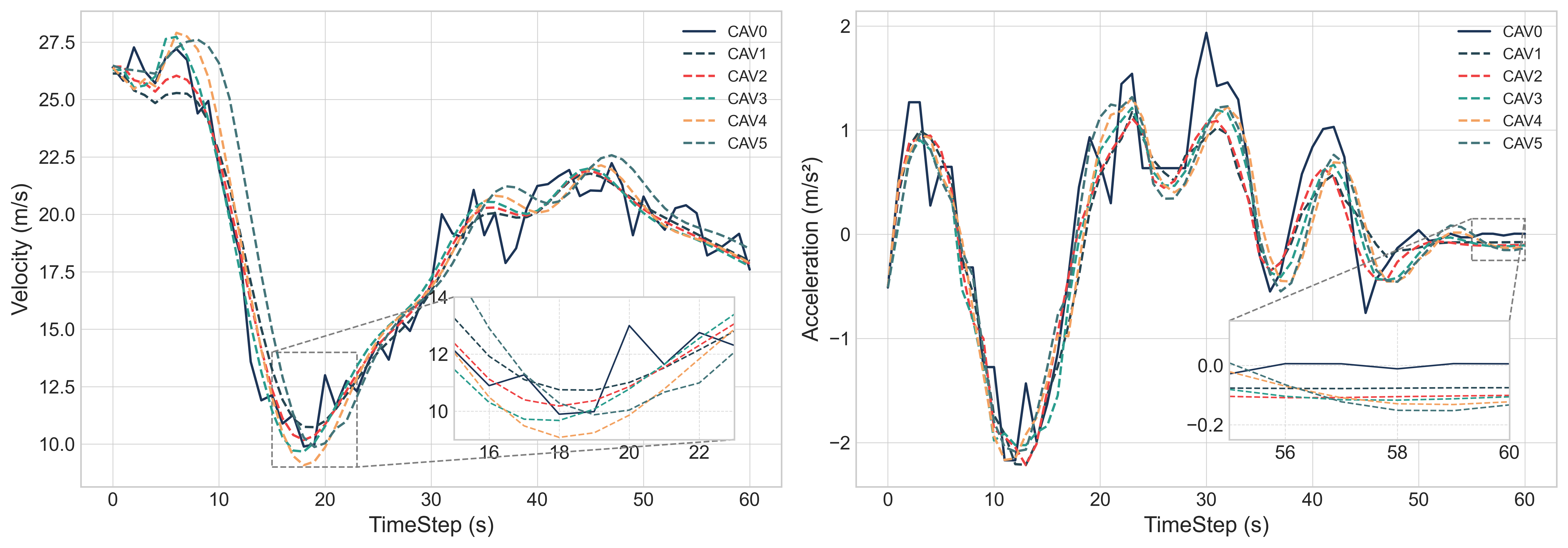}
	}
\subfigure[$D^{3}PG$ Policy]{
	\includegraphics[width=0.850\textwidth]{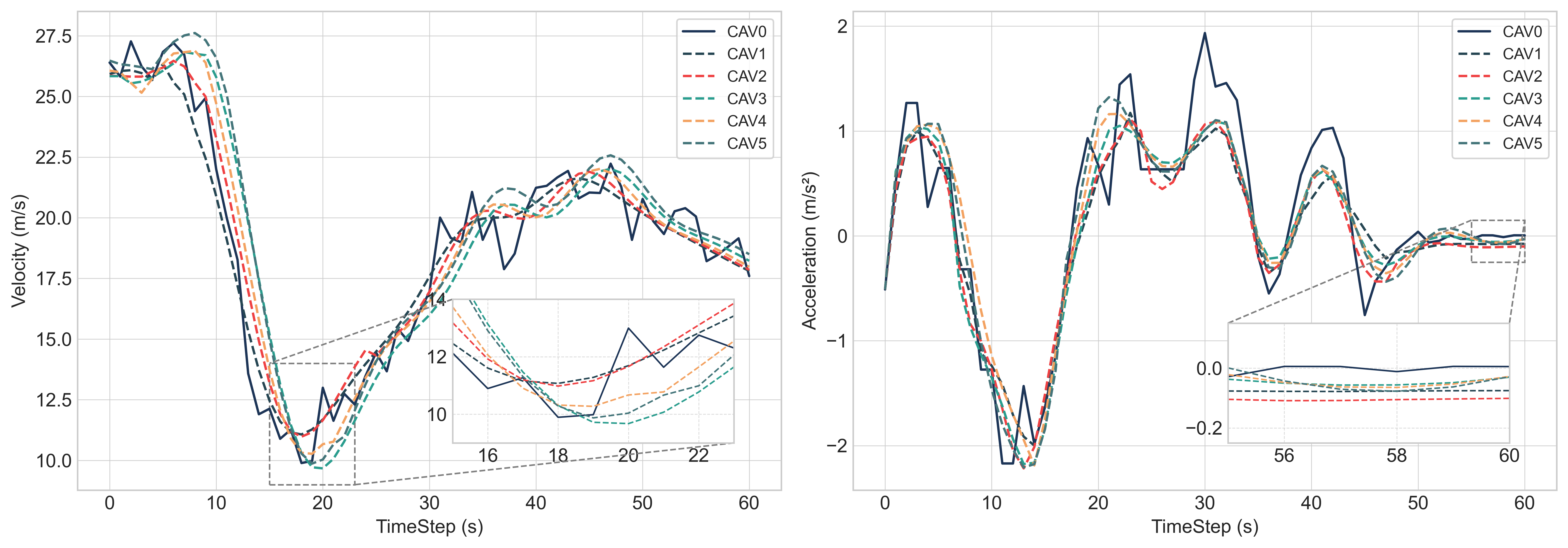}
		}
\subfigure[DCT-MARL Policy]{
	\includegraphics[width=0.850\textwidth]{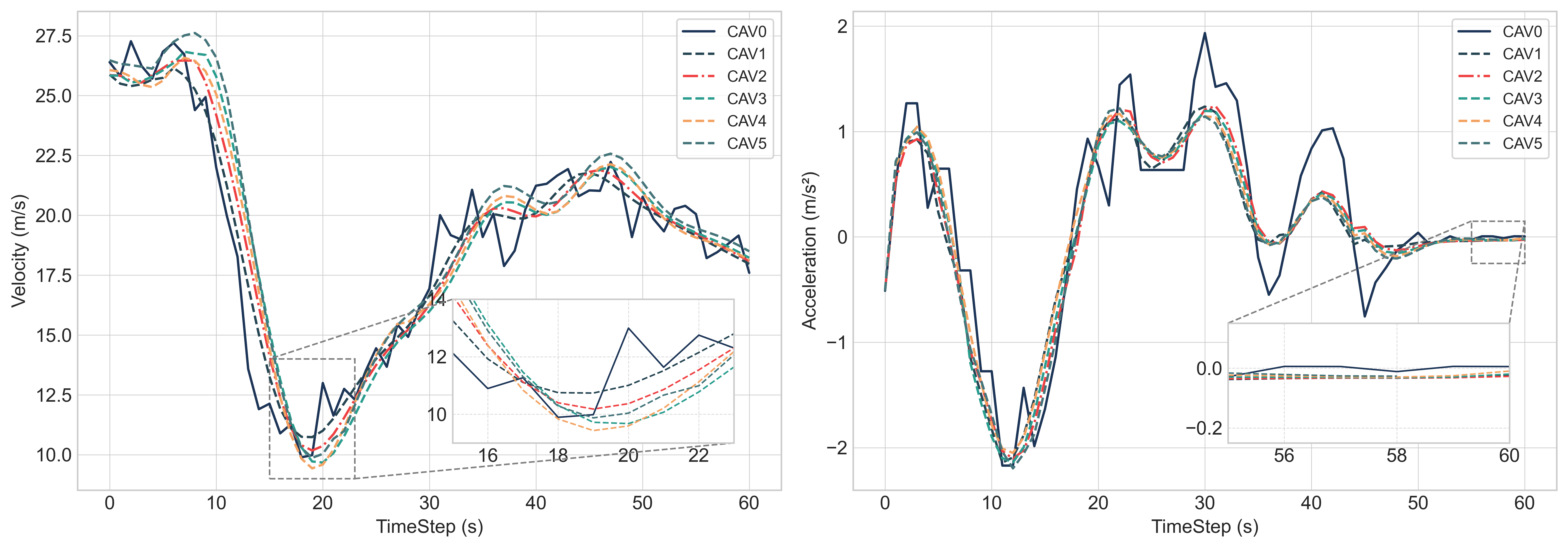}
		}
	\caption{Performance Results of Vehicle Platoon Under Different Strategies.(a) DWOC (b) $D^{3}PG$ (c) DCT-MARL }
\label{fig:vel}
\end{figure*}

Figure~\ref{fig:vel} illustrates the temporal responses of vehicle velocity and acceleration under an oscillatory driving scenario. Across all methods, a noticeable attenuation of acceleration fluctuations is observed along the platoon, indicating partial suppression of error propagation. However, notable discrepancies persist in the response to abrupt velocity variations and in the smooth attenuation of acceleration during convergence.

The DWOC strategy (Fig.\ref{fig:vel} (a)) achieves control stability by adaptively re-weighting front-vehicle information, offering a degree of robustness to delay and packet loss. However, noticeable acceleration spikes between 15s and 25s reveal challenges in ensuring smooth transitions during sudden lead vehicle maneuvers. In contrast, $D^{3}PG$ (Fig.\ref{fig:vel}(b)) improves inter-agent coordination by selecting communication neighbors using attention-based mechanisms. Nonetheless, due to its single-agent DRL design lacking full MARL coordination, downstream vehicles show delayed reactions to upstream disturbances. The proposed DCT-MARL method (Fig.\ref{fig:vel}(c)) consistently delivers the most stable and coordinated control responses, featuring minimal velocity fluctuations and enhanced inter-vehicle coherence, thereby demonstrating superior robustness under non-ideal communication conditions.

Table~\ref{tab:qua} summarizes quantitative results. Both DWOC and $D^3PG$ maintain reasonable performance across the platoon, yet DWOC exhibits disturbance amplification toward the rear of the platoon, indicating limited suppression capability. Although $D^3PG$ alleviates this issue via communication-aware strategies, intermittent oscillations still occur under uncertain  communication conditions. In contrast, DCT-MARL achieves the lowest string stability metrics across all vehicles, particularly in $CAV_3$ to $CAV_5$, demonstrating its superior ability to suppress disturbance propagation. Moreover, DCT-MARL yields the highest overall driving comfort scores, maintaining low acceleration variance even for the last vehicle in the platoon, which is typically most susceptible to disturbance, thereby ensuring an improved passenger experience.

\begin{figure*}[htbp]
	\centering
	\includegraphics[width=0.850\textwidth]{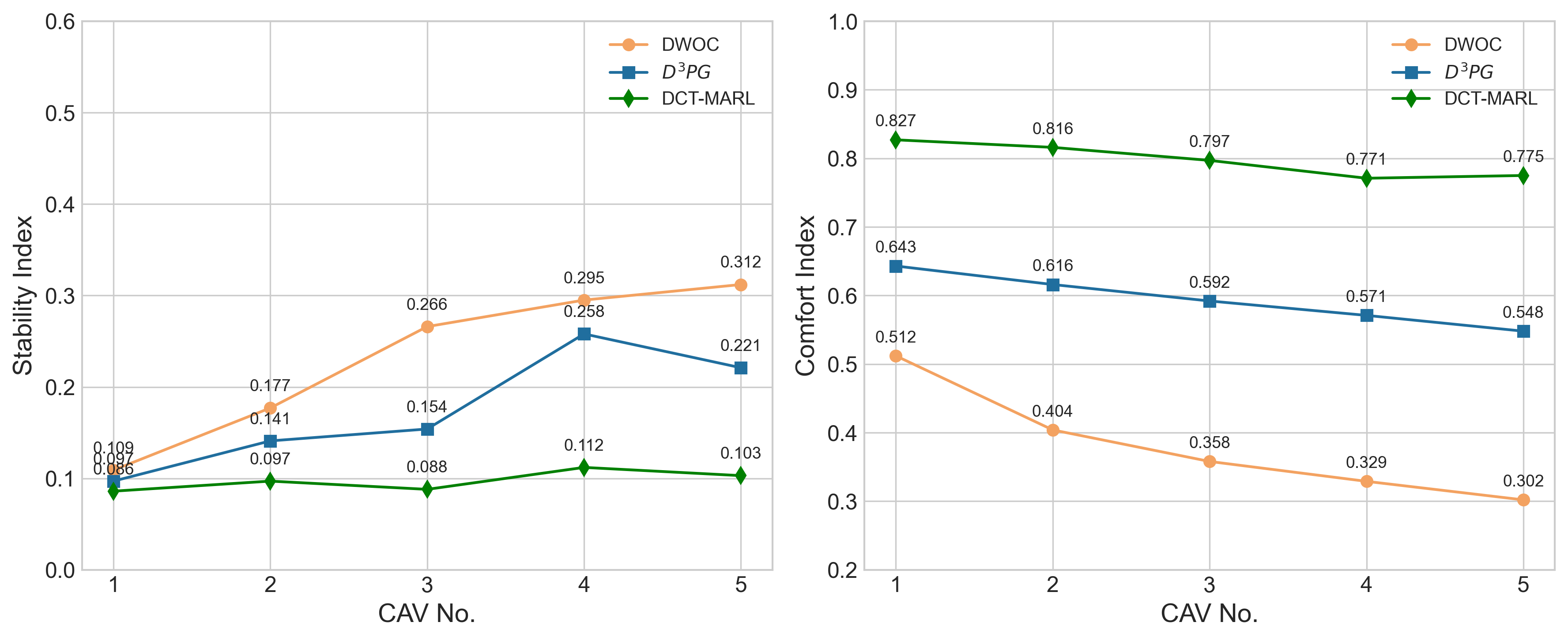}
	\caption{Quantified Metrics of Different Models.}
	\label{fig:com}
\end{figure*}

\begin{table*}[htbp]
	\centering
	\caption{Quantified Metrics for Each CAV in the Platoon Controlled by Different Policys.}
	\label{tab:qua}
	\begin{tabular}{ccccccc}
		\toprule
		&& $CAV_1$  & $CAV_2$ & $CAV_3$ & $CAV_4$ & $CAV_5$ \\
		\midrule
		\multirow{4}{*}{String Stability} 	
		& DWOC         & 0.109 & 0.177 & 0.266 & 0.295 & 0.312 \\
		& $D^{3}PG$    & 0.097 & 0.141 & 0.154 & 0.258 & 0.221 \\
		& \textbf{DCT-MARL}   & \textbf{0.086} & \textbf{0.097} & \textbf{0.088} & \textbf{0.112} & \textbf{0.103} \\
		\midrule
		\midrule
		\multirow{4}{*}{Driving Comfort}  
		& DWOC         & 0.512 & 0.404 & 0.358 & 0.329 & 0.302 \\
		& $D^{3}PG$    & 0.643 & 0.616 & 0.592 & 0.571 & 0.548 \\
		& \textbf{DCT-MARL}   & \textbf{0.827} & \textbf{0.816} & \textbf{0.797} & \textbf{0.771} & \textbf{0.775} \\
		\bottomrule
		\bottomrule
	\end{tabular}
\end{table*}
To illustrate this more clearly, Fig.\ref{fig:com} presents the performance comparison of various control strategies. The DCT-MARL curve maintains a relatively flat comfort trajectory, while all other baselines show progressively deteriorating performance over time. This highlights DCT-MARL’s robustness and resilience under dynamic and uncertain communication conditions. By jointly considering topology adaptation, delay compensation, and cooperative policy learning, the DCT-MARL method achieves superior multi-objective optimization, balancing safety, comfort, and stability more effectively than DWOC, and $D^{3}PG$.

In summary, the proposed DCT-MARL framework demonstrates significant advantages over existing platoon control methods by effectively mitigating the adverse effects of communication delay and packet loss. It consistently enhances string stability and driving comfort throughout the platoon, especially for vehicles more vulnerable to disturbances.

\subsection{Control Performance against Existing MARL Algorithms}
To further assess the advantages of the proposed DCT-MARL framework, this paper compare it against two representative MARL algorithms: Delay-Compensated MARL (DEC-MARL) \cite{wang2024multi}, which explicitly addresses communication delay by extending the state space to incorporate historical control outputs under a fixed communication topology. Learning-based MARL (LEB-MARL) \cite{ding2020learning}, which leverages causal inference to infer communication topologies and regularizes agent policies to enhance performance in multi-agent cooperative scenarios.

\begin{figure}[htbp]
	\centering
	\includegraphics[width=0.50\textwidth]{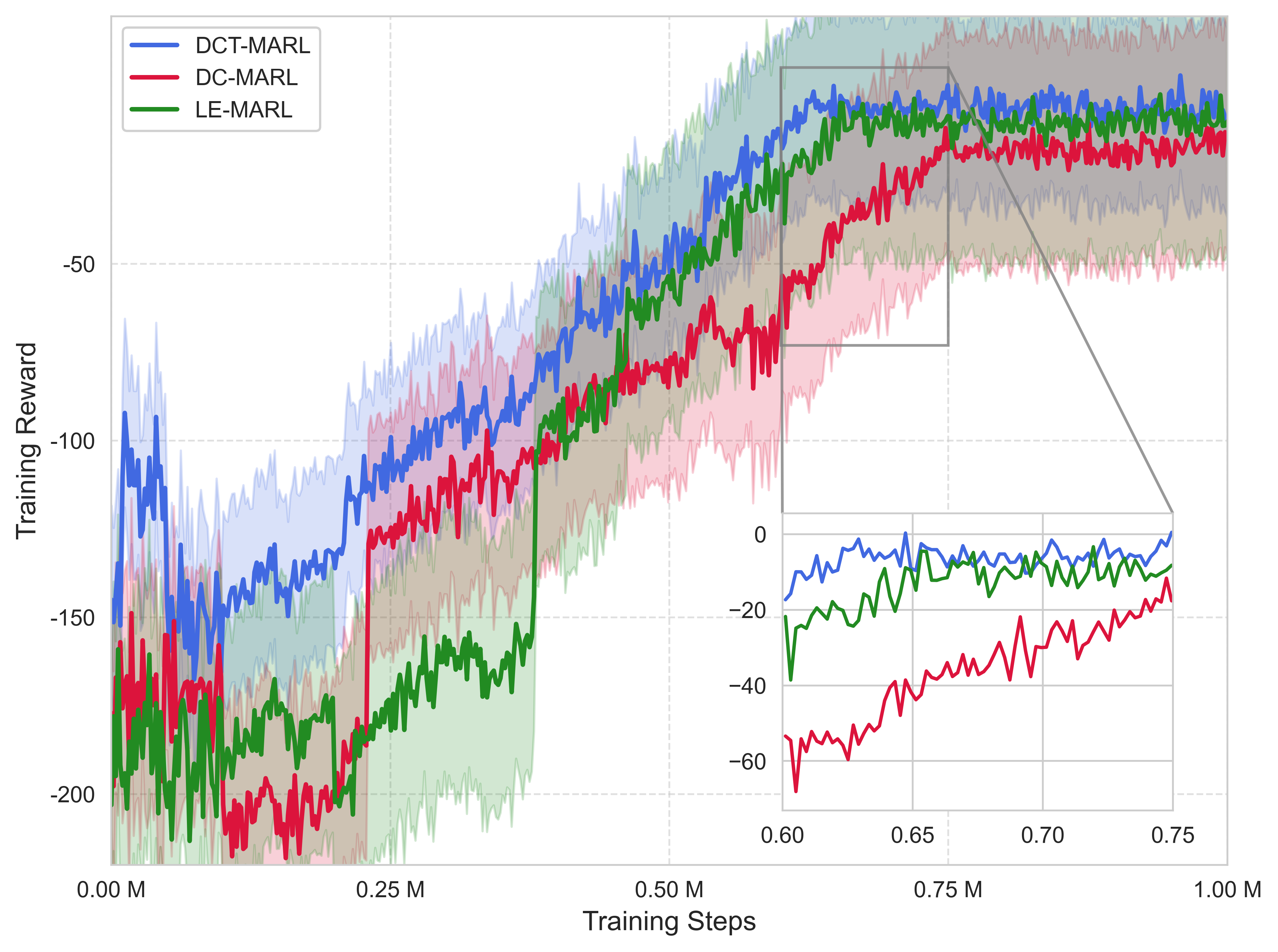}
	\caption{Average Episode Reward vs. Training Steps of Various Methods.}
	\label{fig:reward}
\end{figure}

To assess learning efficiency, the average episodic return is depicted in Fig. \ref{fig:reward}, which compares the performance of the  proposed DCT-MARL algorithm against other MARL methods. As training progresses, DCT-MARL consistently achieves higher rewards, indicating more effective policy optimization. In contrast, DEC-MARL and LEB-MARL exhibit slower convergence and lower final reward values, reflecting reduced learning efficiency in dynamic communication environments. These results demonstrate that DCT-MARL outperforms the baselines in both convergence speed and reward performance, highlighting its superior adaptability and learning capability.

Table \ref{tab:topo} compares the control performance of the three methods in terms of string stability and driving comfort for each vehicle within the platoon. While DEC-MARL enhances system stability to a certain extent under non-ideal communication conditions, the reliance on a fixed communication topology and absence of dynamic adjustment mechanisms cause significant performance degradation under high packet loss rates and dynamic communication structures. This results in greater spacing fluctuations among following vehicles and a noticeable decline in driving comfort. LEB-MARL incorporates agent correlation and demonstrates certain advantages in control accuracy. However, it fails to account for the time-varying nature of communication quality, limiting robustness under packet loss and topological disturbances and leading to more pronounced oscillations in string stability. In contrast, DCT-MARL consistently delivers more robust and stable control outcomes across all CAVs in the platoon.

\begin{table*}[ht]
	\centering
	\caption{Quantified Metrics for Each CAV in the Platoon Controlled by FCT-MARL and DCT-MARL.}
	\label{tab:topo}
	\begin{tabular}{ccccccc}
		\toprule
		&& $CAV_1$  & $CAV_2$ & $CAV_3$ & $CAV_4$ & $CAV_5$ \\
		\midrule
		\multirow{2}{*}{String Stability} 	
		& DEC-MARL     & 0.114 & 0.138 & 0.167 & 0.248 & 0.279 \\
		& LEB-MARL  & 0.103 & 0.125 & 0.134 & 0.197 & 0.221 \\ 
		& \textbf{DCT-MARL}   & \textbf{0.086} & \textbf{0.097} & \textbf{0.088} & \textbf{0.112} & \textbf{0.103} \\
		\midrule
		\multirow{2}{*}{Driving Comfort }  
		& DEC-MARL       & 0.681 & 0.628 & 0.538 & 0.516 & 0.505 \\
		& LEB-MARL  & 0.739 & 0.712 & 0.6358 & 0.673 & 0.629 \\
		& \textbf{DCT-MARL}   & \textbf{0.827} & \textbf{0.816} & \textbf{0.797} & \textbf{0.771} & \textbf{0.775} \\
		\bottomrule
		\bottomrule
	\end{tabular}
\end{table*}

Beyond training performance, the runtime efficiency of the proposed DCT-MARL framework is evaluated during testing. On a local platform equipped with a 2.0 GHz quad-core Intel Core i5 processor and 16 GB of 3733 MHz LPDDR4X memory, executing a 600-step simulation required approximately 3.7 seconds, indicating that each decision step takes approximately 6 ms. Given the 100 ms sampling period of vehicle control systems, this inference time is well within the control horizon. These results demonstrate that DCT-MARL not only achieves superior control performance but also exhibits high inference efficiency, rendering it suitable for realtime deployment in computationally constrained environments.

\subsection{Eiffectiveness of Dynamic Communictaion Topology}
\begin{figure}[htbp]
	\centering
	\includegraphics[width=0.45\textwidth]{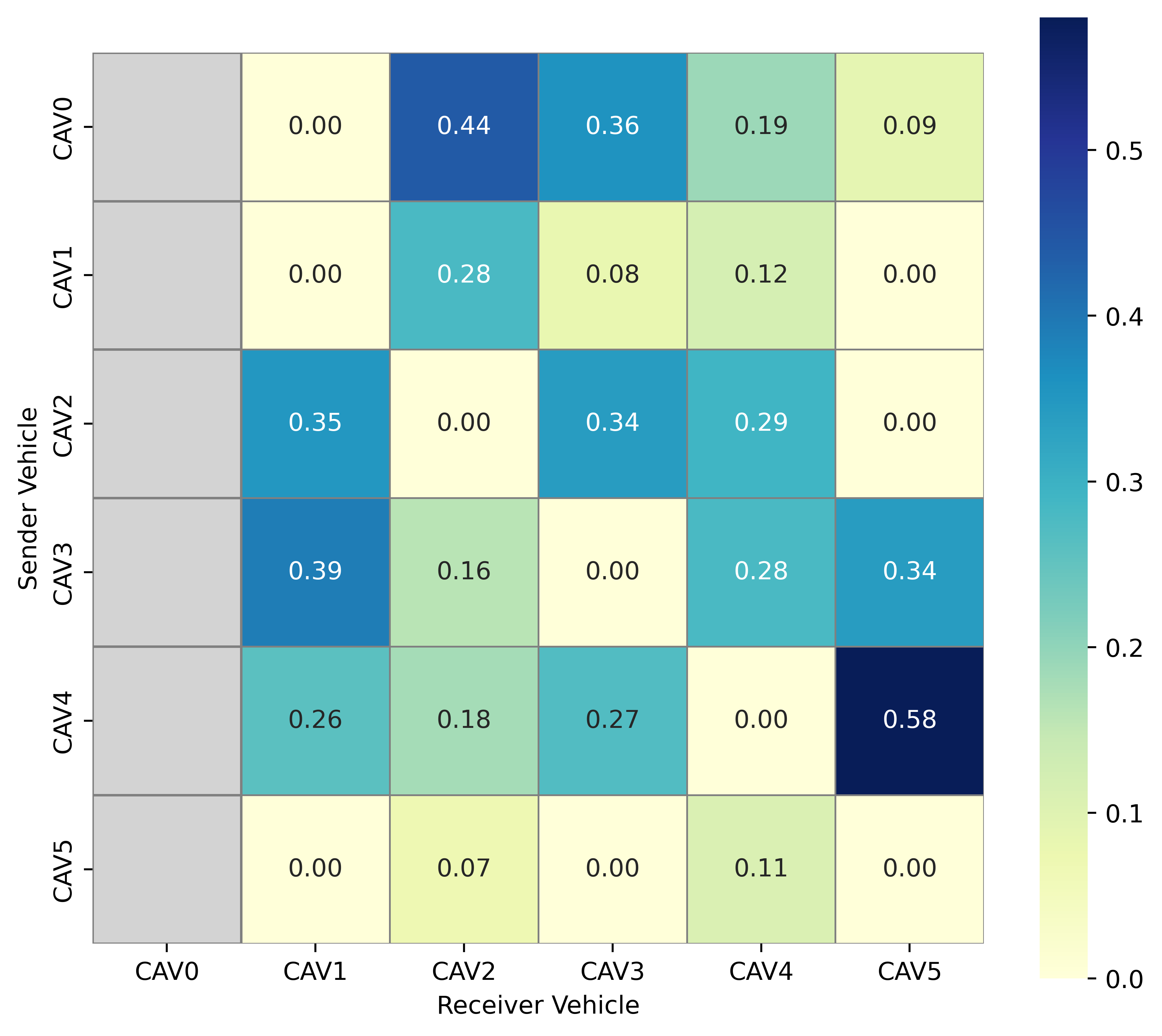}
	\caption{Communication Heatmap of Connected Platoon with Dynamic Topology.}
	\label{fig:hot}
\end{figure}

To further examine the necessity and rationality of the dynamic communication topology mechanism, Fig. \ref{fig:hot} illustrates the communication heatmap among six platoon memebers during the testing phase. 

In the heatmap, the color intensity indicates the frequency of message exchanges between agents. The vertical axis represents the message sender, while the horizontal axis corresponds to the receiver. In this visualization, a darker color indicates a higher likelihood of information exchange, reflecting stronger communication between specific vehicle pairs. This allows us to intuitively assess how communication patterns evolve under the dynamic topology. 

Notably, as the leading vehicle operates as a noncontrollable preset agent, it is only possible to observe whether other vehicles receive messages from it, without direct access to the messages received by the leader itself. Accordingly, the corresponding cells representing the leader as a receiver are shaded in light gray to indicate the absence of observable data. 

It can be observed that while front-vehicle information receives predominant attention in the control process, rear vehicle data also demonstrates significant utilization. This observation is consistent with the findings of [5], highlighting the latent value of rear vehicle states in control decision making. Compared to traditional fixed topologies, the dynamic topology can adaptively adjust communication agents in real time based on control needs and information value. These results demonstrate the meaningfulness and effectiveness of incorporating dynamic communication topology under non-ideal communication conditions.

\section{Conclusions and Future Works} \label{sec:conclusion}
This paper proposes a DCT-MARL framework to address connected vehicle platoon control under non-ideal V2V communications characterized by time-varying delay and packet loss. Specifically, a multi-key gated communication mechanism is introduced, which dynamically adjusts the communication topology based on the correlations among vehicles and their current communication status to effectively mitigate packet loss. Furthermore, a state augmentation scheme is designed to incorporate historical control actions and communication delay information into each agent’s state space, ensuring the preservation of the Markov property despite delayed observations. This enhancement further improves the robustness of the system in realistic V2V communication environment. Experimental results demonstrate that the proposed approach significantly improves platoon stability and driving comfort of vehicle platoon compared to existing advanced methods. Future research will further investigate the mechanism of non-ideal communication and explore the joint optimization of communication and control, enabling more adaptive and resource-efficient platoon coordination.

\section*{Acknowledgments}
This work was supported by Beijing Natural Science Foundation under Grant L202018, the National Natural Science Foundation of China under Grant 61931005, the Key Laboratory of Internet of Vehicle Technical Innovation and Testing (CAICT), Ministry of Industry and Information Technology under Grant No. KL-2023-001.

\bibliographystyle{IEEEtran}

\bibliography{references}

% \balance
\begin{IEEEbiography}
	[{\includegraphics[width=1in,height=1.25in,clip,keepaspectratio]{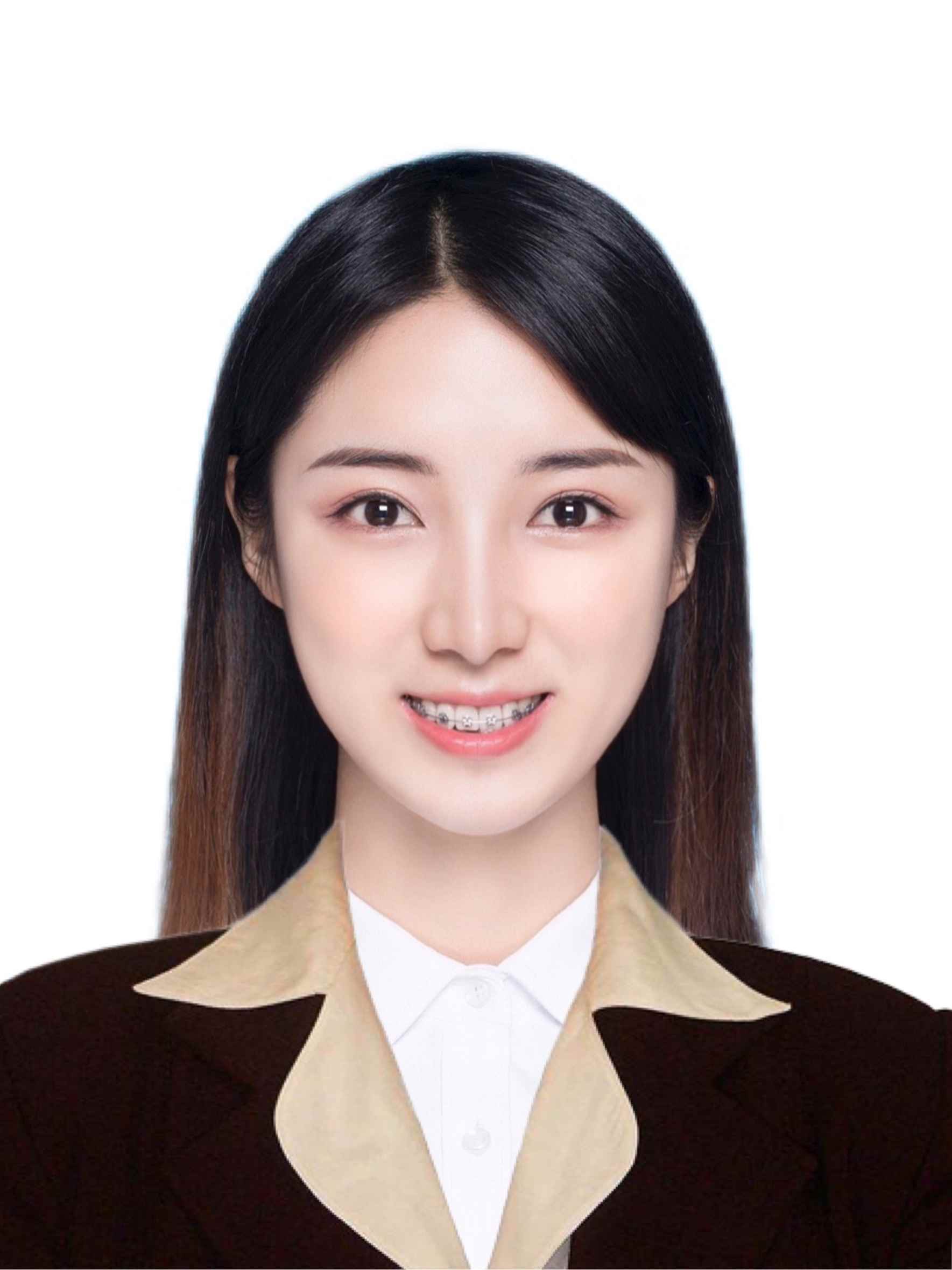}}] 
	{Yaqi Xu} received her M.Sc in computer technology from Jilin University in 2022. She is currently pursuing a Ph.D. degree from State Key Laboratory of Networking and Switching Technology, Beijing University of Posts and Telecommunications (BUPT). She is interested in topics related to vehicular networks, especially Connected and Autonomous Vehicles (CAVs) Platoon and Vehicle to Everything (V2X) Communication. 
	\vspace{-15pt}
\end{IEEEbiography}

\vspace{-4mm}

\begin{IEEEbiography}
	[{\includegraphics[width=1in,height=1.25in,clip,keepaspectratio]{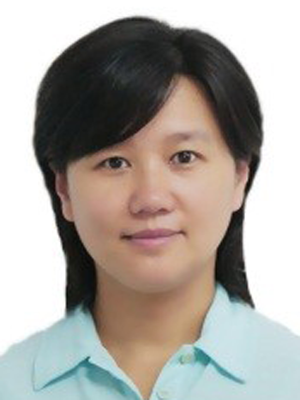}}] 
	{Yan Shi} received her Ph.D. degree from BUPT in 2007. She is currently an associated professor of school of computer science (national pilot software engineering school) and State Key Laboratory of Networking and Switching Technology, BUPT. Her current research interests include vehicular networks and applications, as well as mobile computing.
	\vspace{-15pt}
\end{IEEEbiography}

\vspace{-4mm}

\begin{IEEEbiography}
	[{\includegraphics[width=1in,height=1.25in,clip,keepaspectratio]{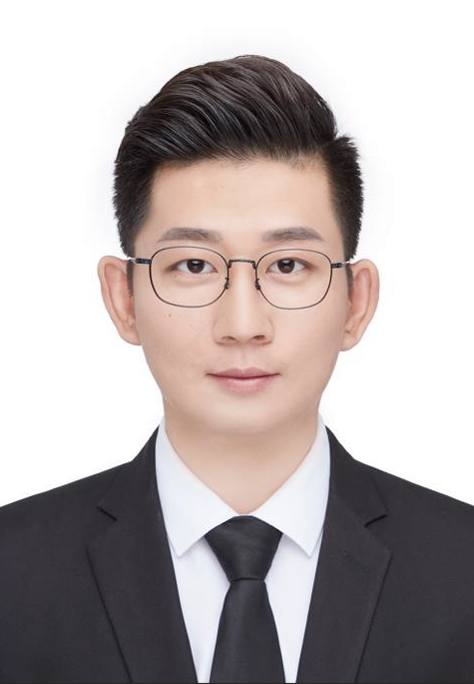}}] 
	{Jin Tian} is currently pursuing the Ph.D. degree from the State Key Laboratory of Networking and Switching Technology, BUPT. His research interests include Cellular Vehicle-to-Everything (C-V2X) and the application of machine learning in vehicular networks.
	\vspace{-15pt}
\end{IEEEbiography}

\vspace{-4mm}

\begin{IEEEbiography}
	[{\includegraphics[width=1in,height=1.25in,clip,keepaspectratio]{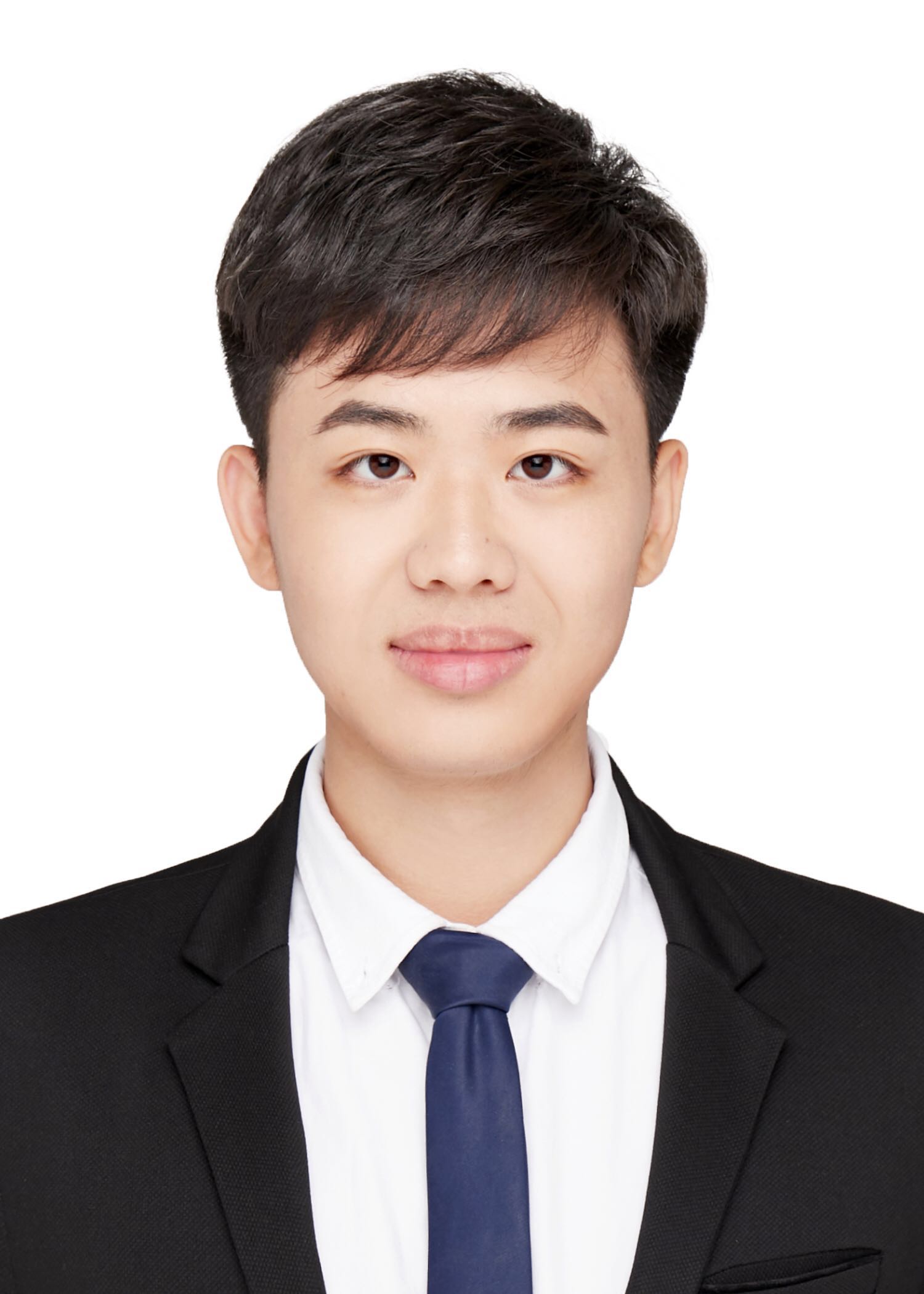}}] 
	{Fanzeng Xia} received: i) a B.A.Sc. degree in Computer Engineering from Queen's University, Kingston, Canada, in 2019; ii) a B.E. degree in Software Engineering from Jilin University, Changchun, China, in 2019; iii) an M.Sc. degree in Computer Science from New York University, New York, United States, in 2021. He is currently a Ph.D. candidate at The Chinese University of Hong Kong, Shenzhen, China, advised by Professor Tongxin Li. His research focuses on foundation models, reinforcement learning, and in-context decision-making.
	\vspace{-15pt}
\end{IEEEbiography}

\vspace{-4mm}

\begin{IEEEbiography}[{\includegraphics[width=1in,height=1.25in,clip,keepaspectratio]{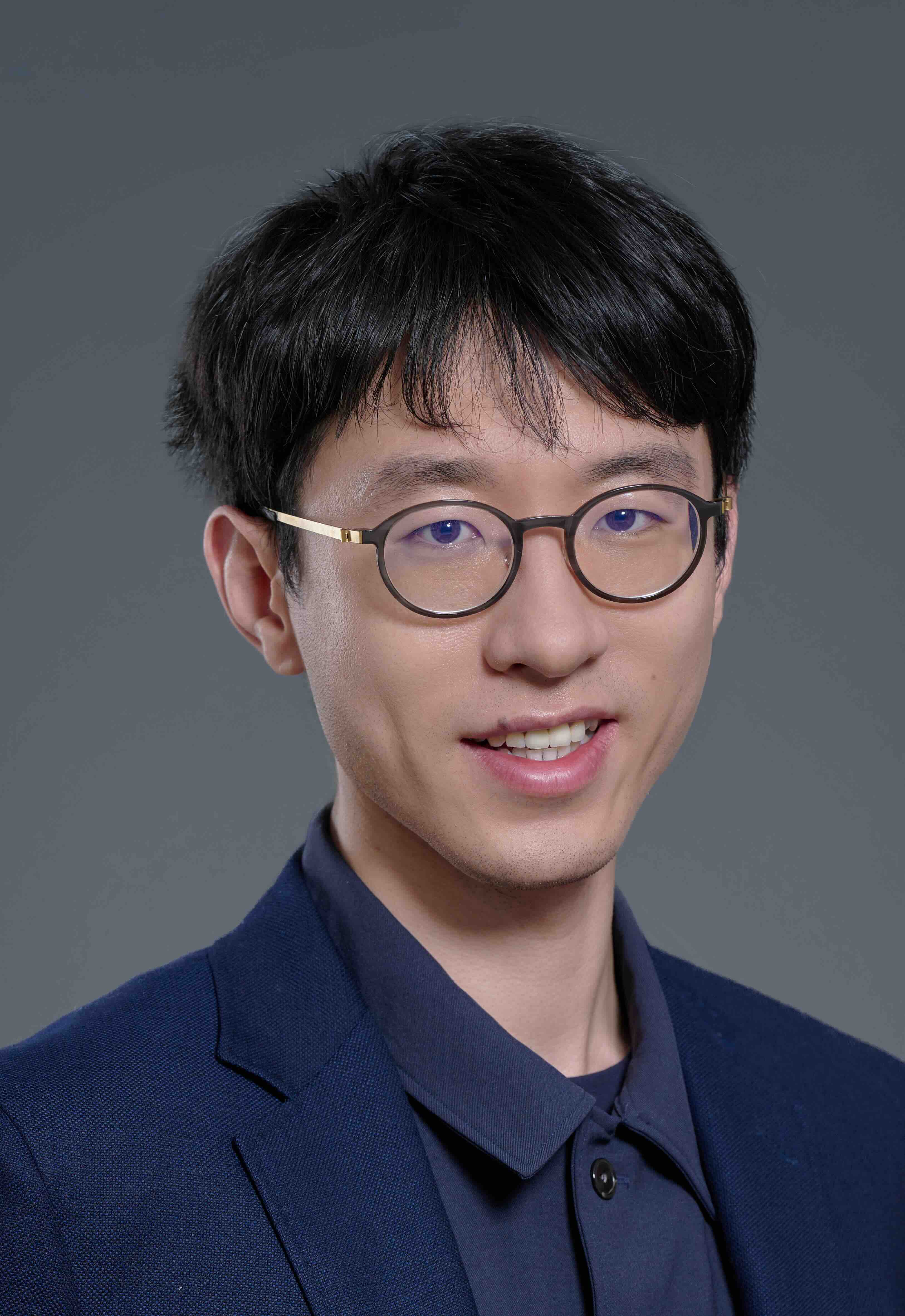}}]
	{Tongxin Li}{\space} received his Ph.D. degree from Caltech in the Department of Computing and Mathematical Sciences in 2022. He is an Assistant Professor, Presidential Young Fellow in the School of Data Science (SDS), CUHK-Shenzhen and an Adjunct Assistant Professor at Shenzhen Loop Area Institute. He interned as an applied scientist at Amazon Web Services (AWS) in 2020 and 2021. His research interests focus on interdisciplinary topics in learning, control, and cyber-physical systems. He was a recipient of the 2022 ACM SIGEnergy Doctoral Dissertation Award (Honorable Mention).
\end{IEEEbiography}

\vspace{-4mm}

\begin{IEEEbiography}
	[{\includegraphics[width=1in,height=1.25in,clip,keepaspectratio]{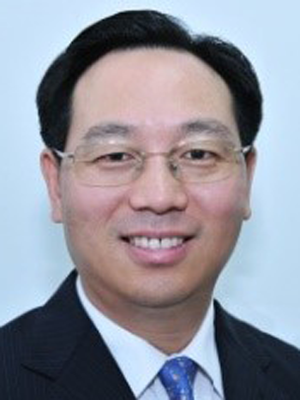}}] 
	{Shanzhi Chen} received the bachelor’s degree from Xidian University, Xi’an, China in 1991, and the Ph.D. degree from Beijing University of Posts and Telecommunications, Beijing, China in 1997. He joined the Datang Telecom Technology and Industry Group and the China Academy of Telecommunication Technology (CATT) in 1994 and has been serving as the CTO and EVP of Research and Development from 2008 to 2012. Now, Dr. Chen is the EVP of R$\&$D and CTO of China Information and Communication Technology Group Co., Ltd. (CICT), the director of State Key Laboratory of Wireless Mobile Communications, and the board director of SMIC. He has contributed to the design, standardization, and development of 4G TD-LTE, 5G, and C-V2X communication systems. His current research interests include B5G/6G, C-V2X, and integrated terrestrial-satellite communication systems. He is an IEEE fellow, has served as the Editor-in-Chief of Telecommunications Science, an Area Editor of the IEEE Internet of Things Journal, and an Editor of IEEE Network, as well as a member of TPC Chairs of many international conferences.
	\vspace{-15pt}
\end{IEEEbiography}

\vspace{-4mm}

\begin{IEEEbiography}
	[{\includegraphics[width=1in,height=1.25in,clip,keepaspectratio]{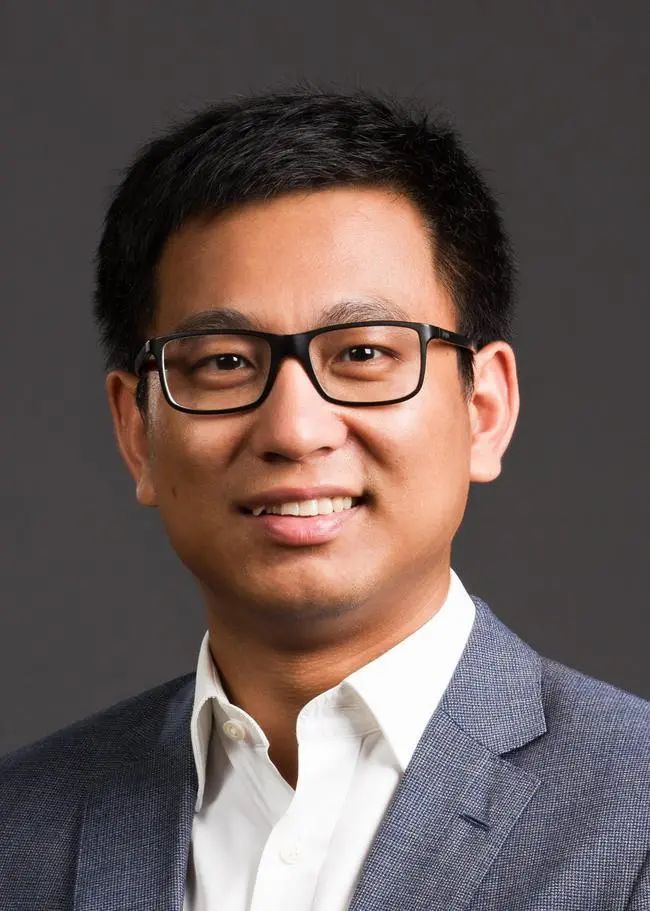}}] 
	{Yuming Ge} received his Ph.D. degree from Institute of Computing Technology, Chinese Academy of Sciences in 2013. He is the director of Automotive and Transportation Department of China Academy of Information and Communications Technology (CAICT) and the leader of C-V2X Working Group of IMT-2020 (5G) Promotion Group. He is interested in topics related to C-V2X communication and vehicular applications.
	\vspace{-15pt}
\end{IEEEbiography}

\end{document}